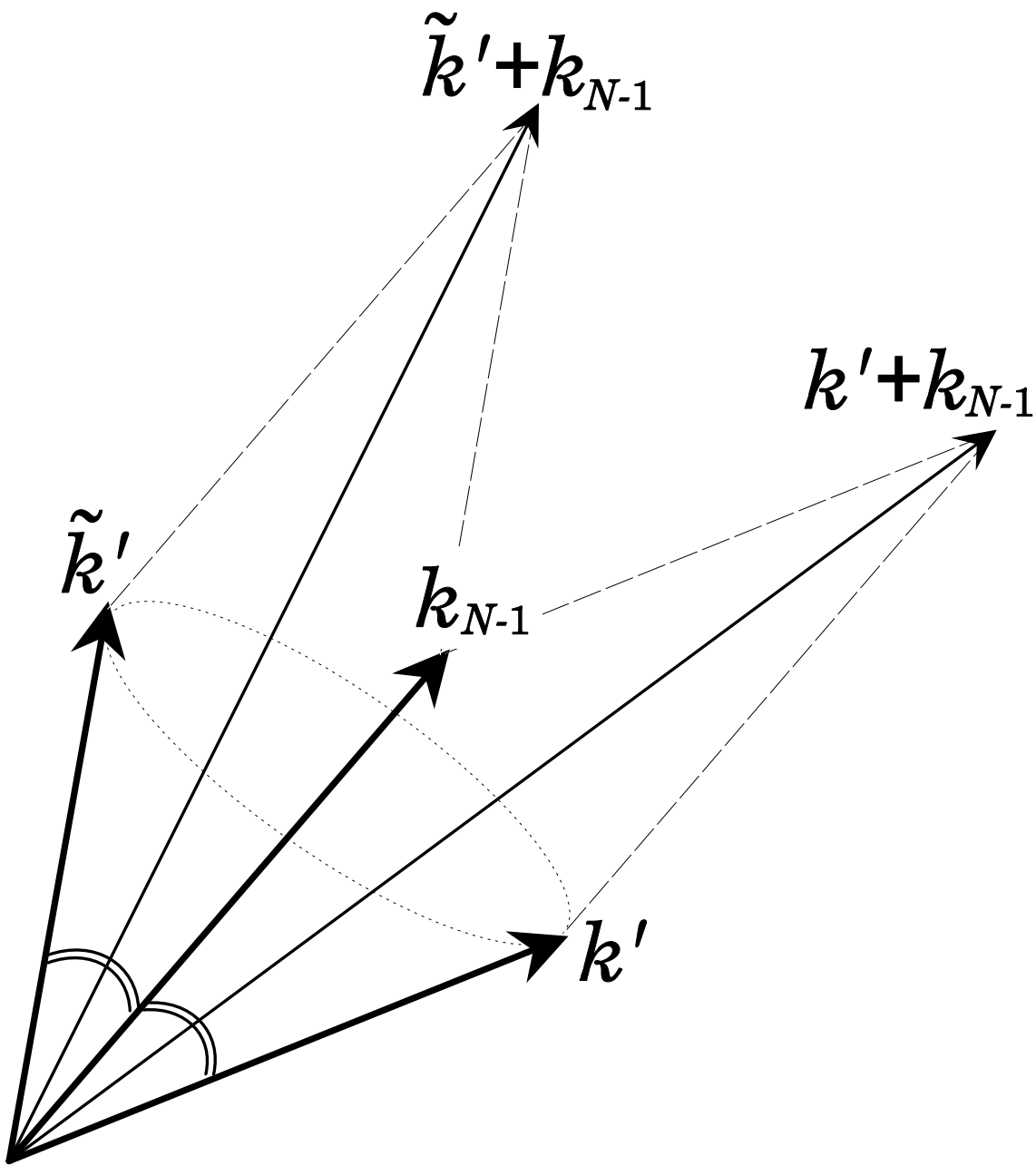

**Fig.1**

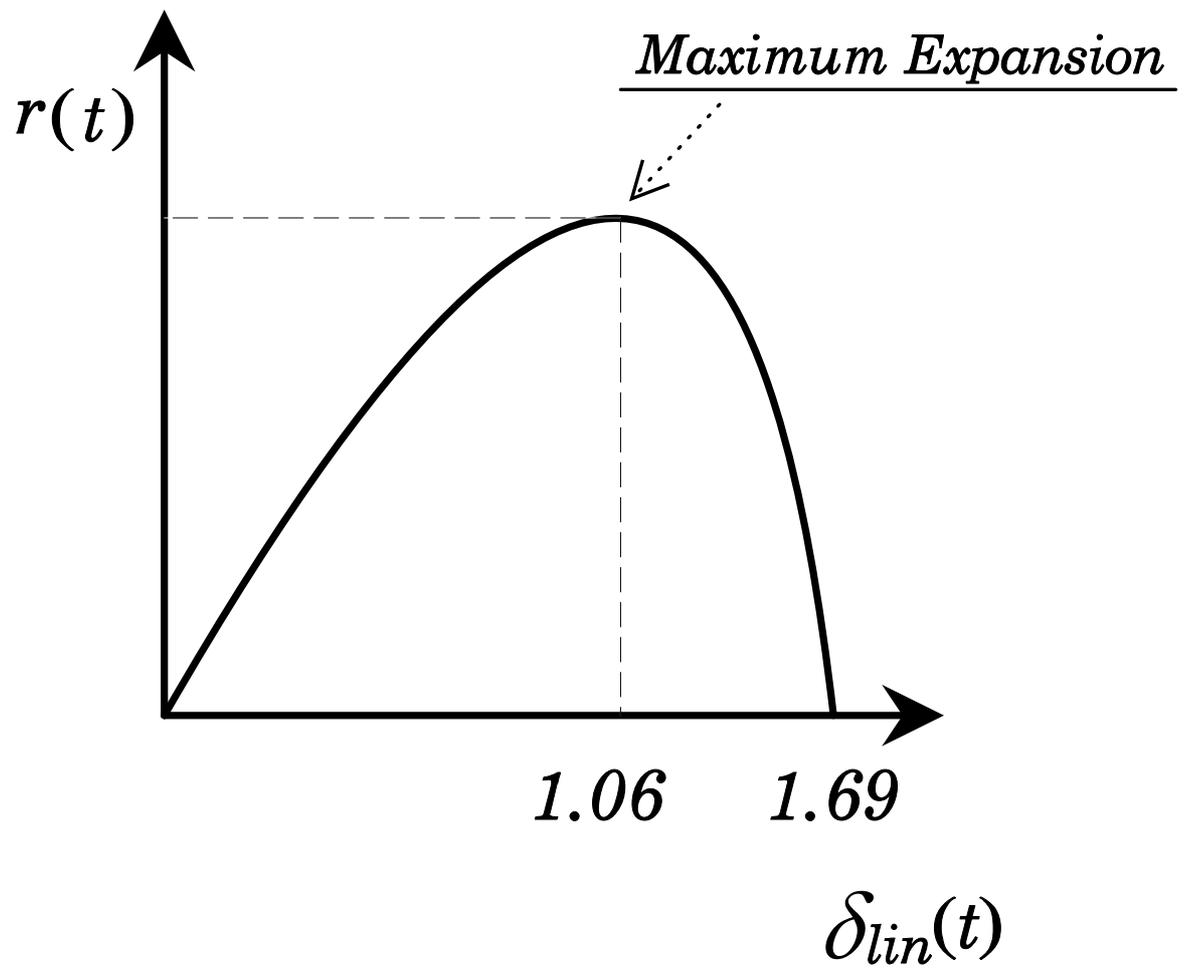

**Fig. 2**

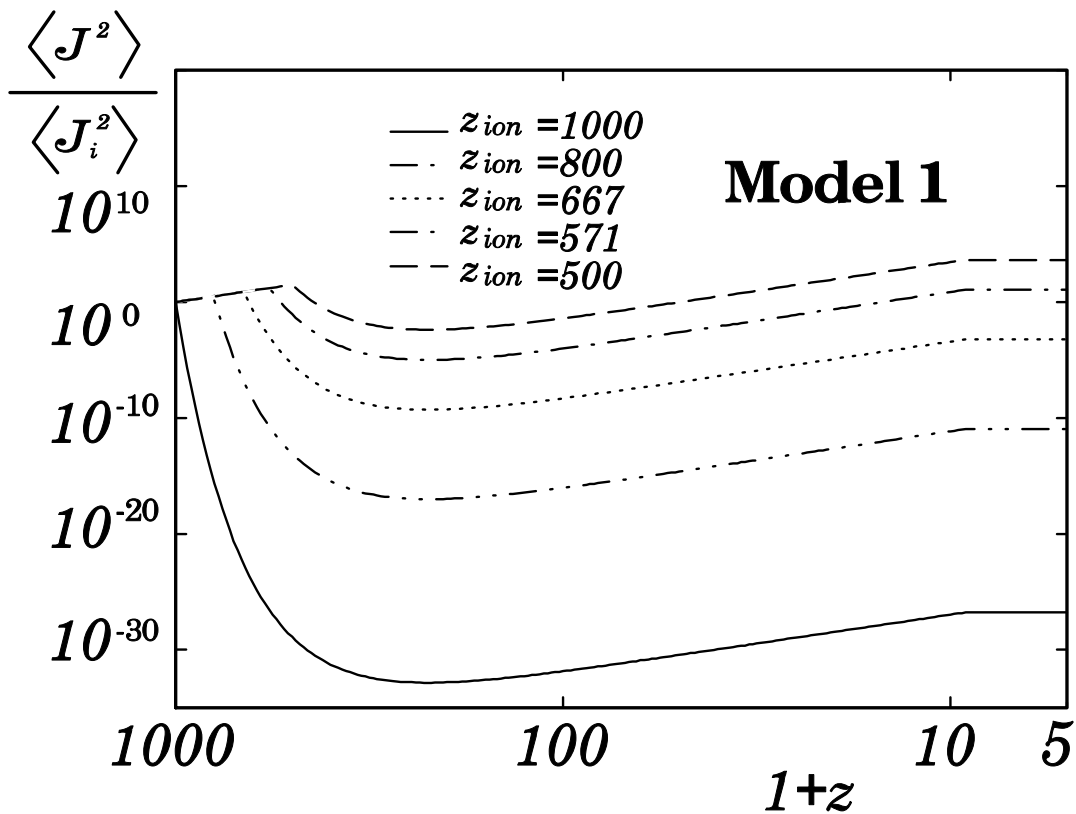

Fig.3a

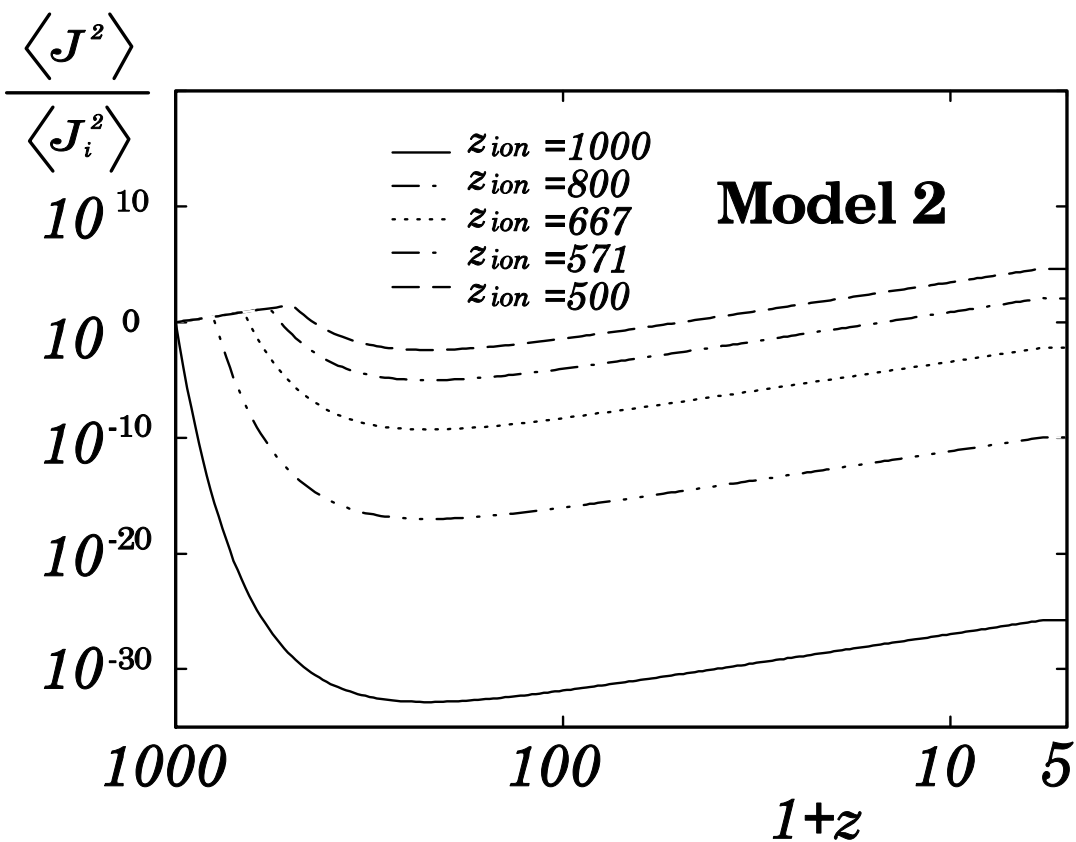

Fig.3b

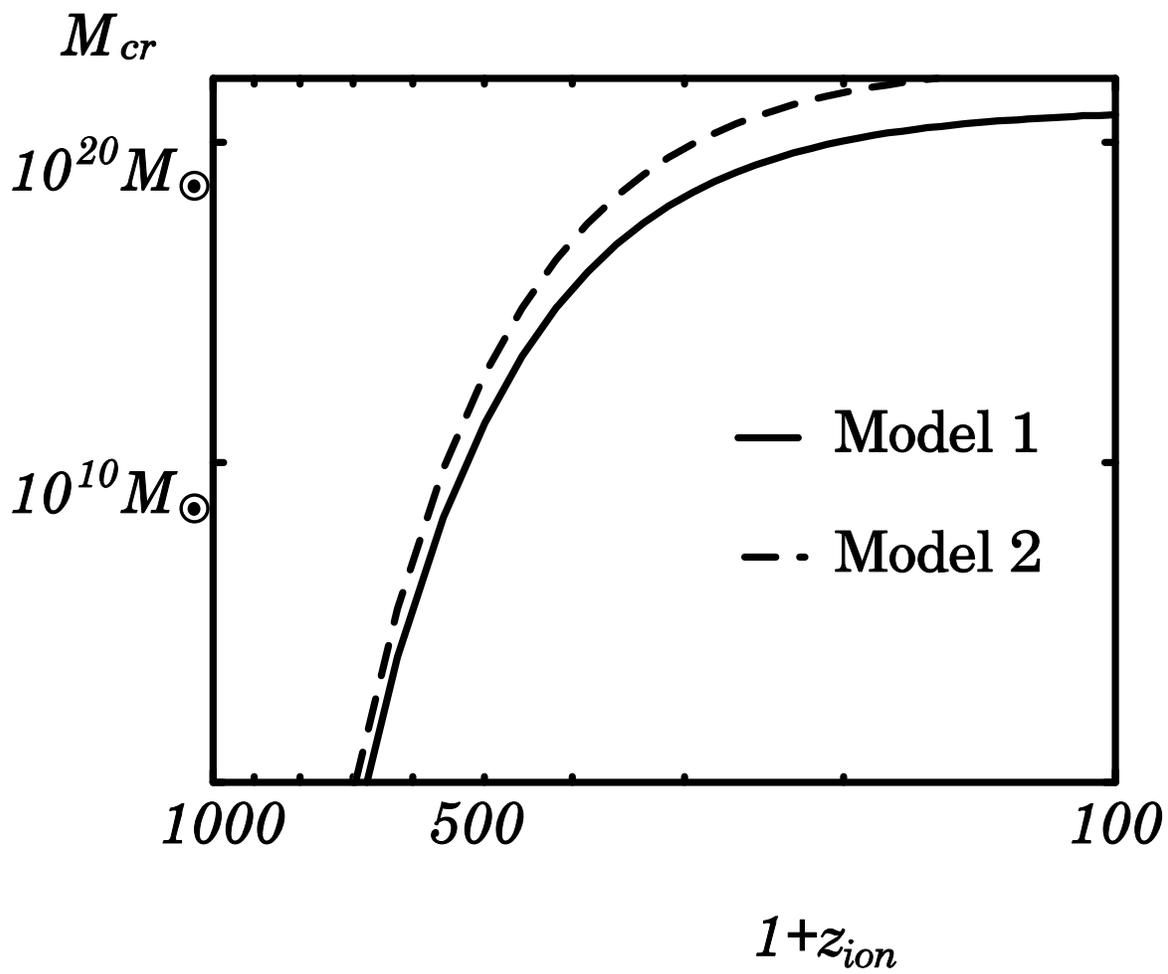

**Fig. 4**

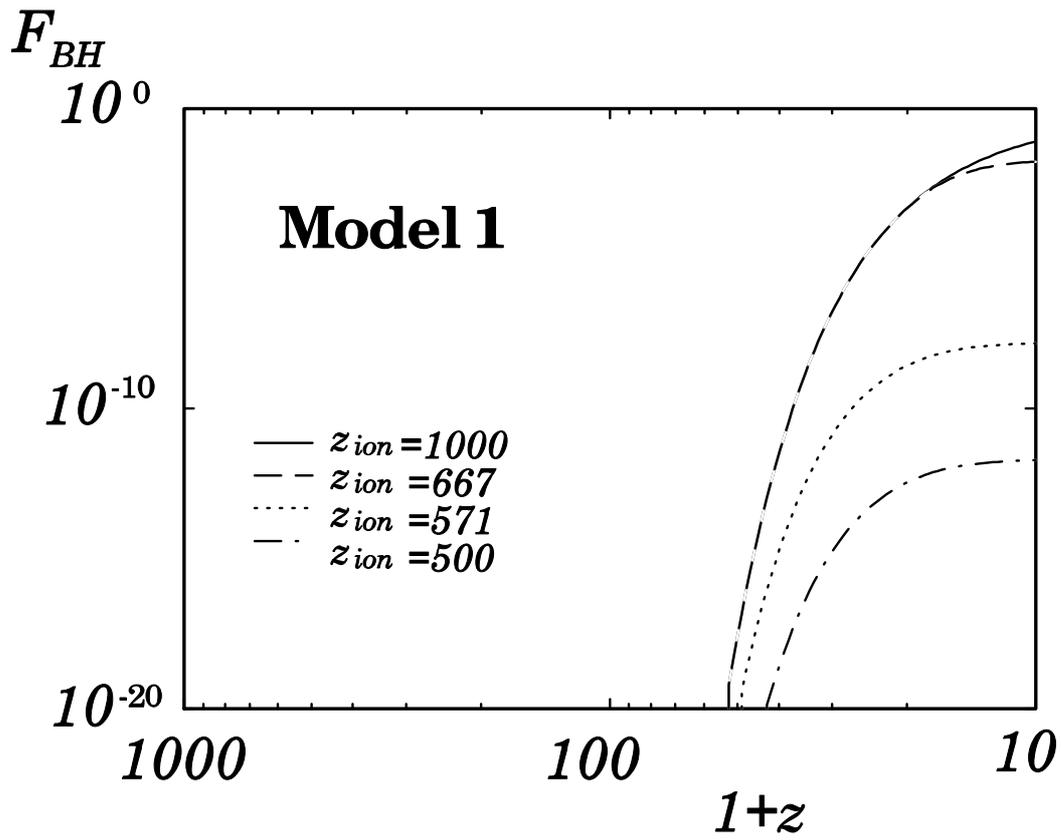
Fig.5a

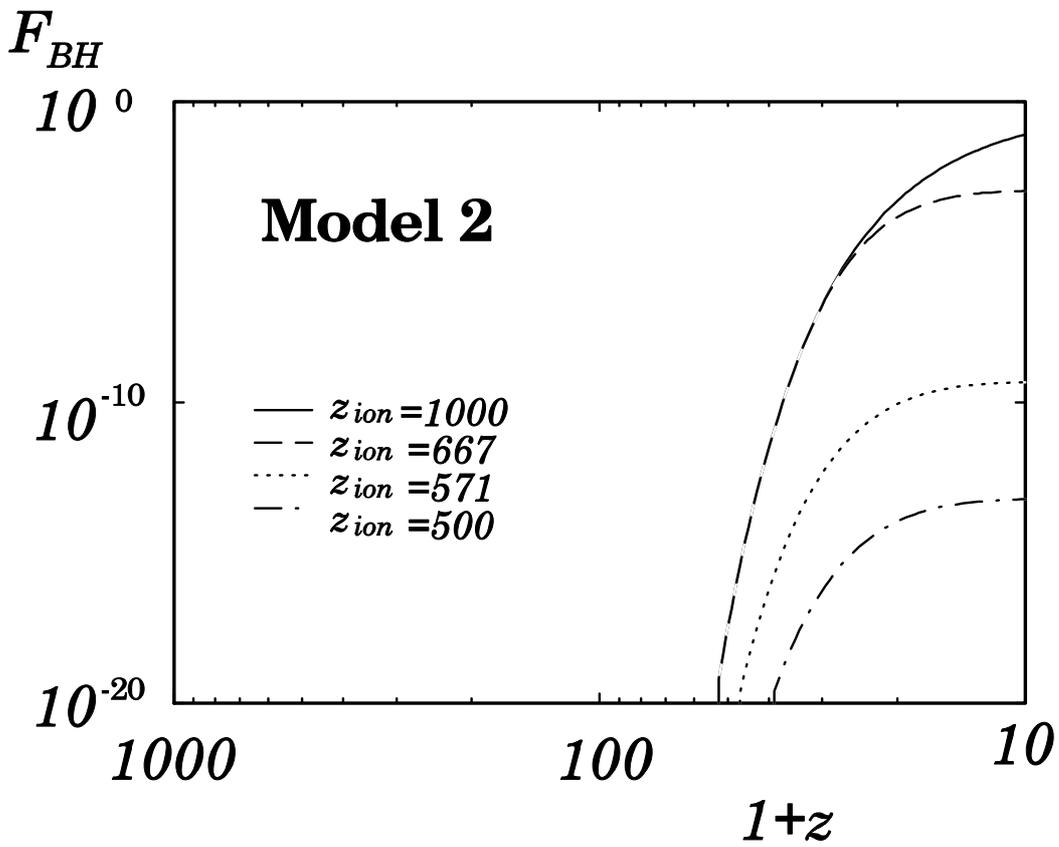
Fig.5b

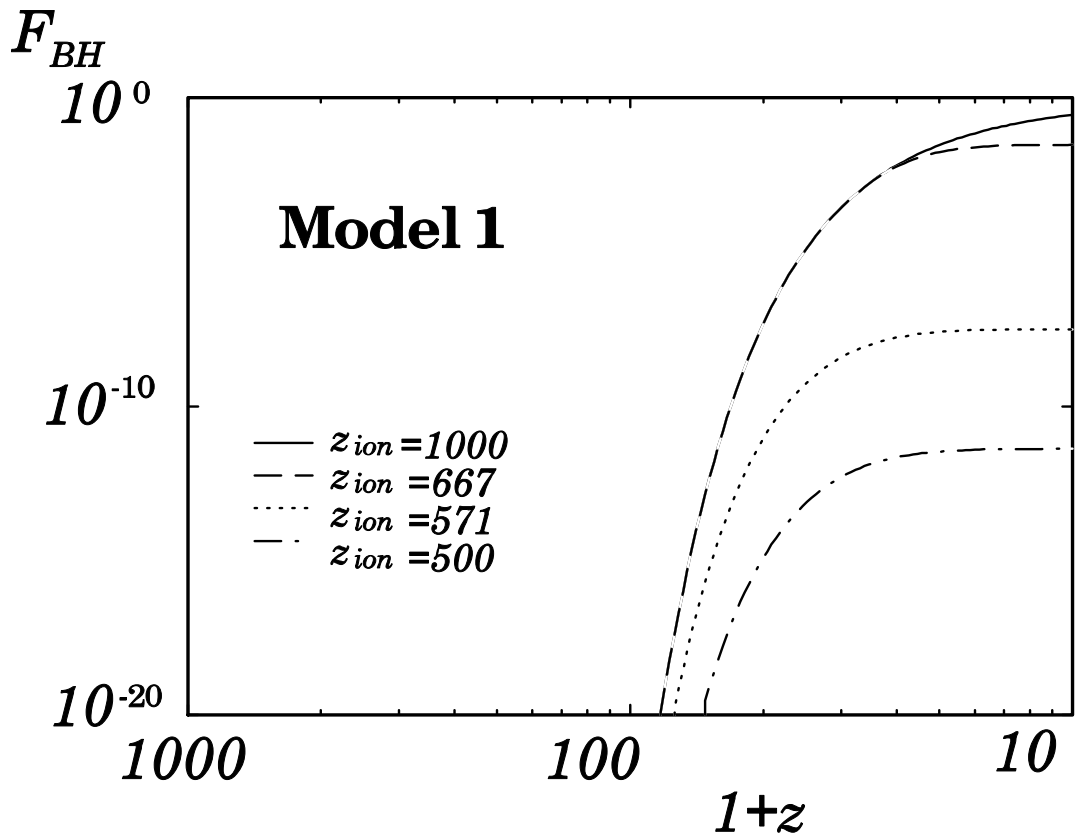
Fig.6a

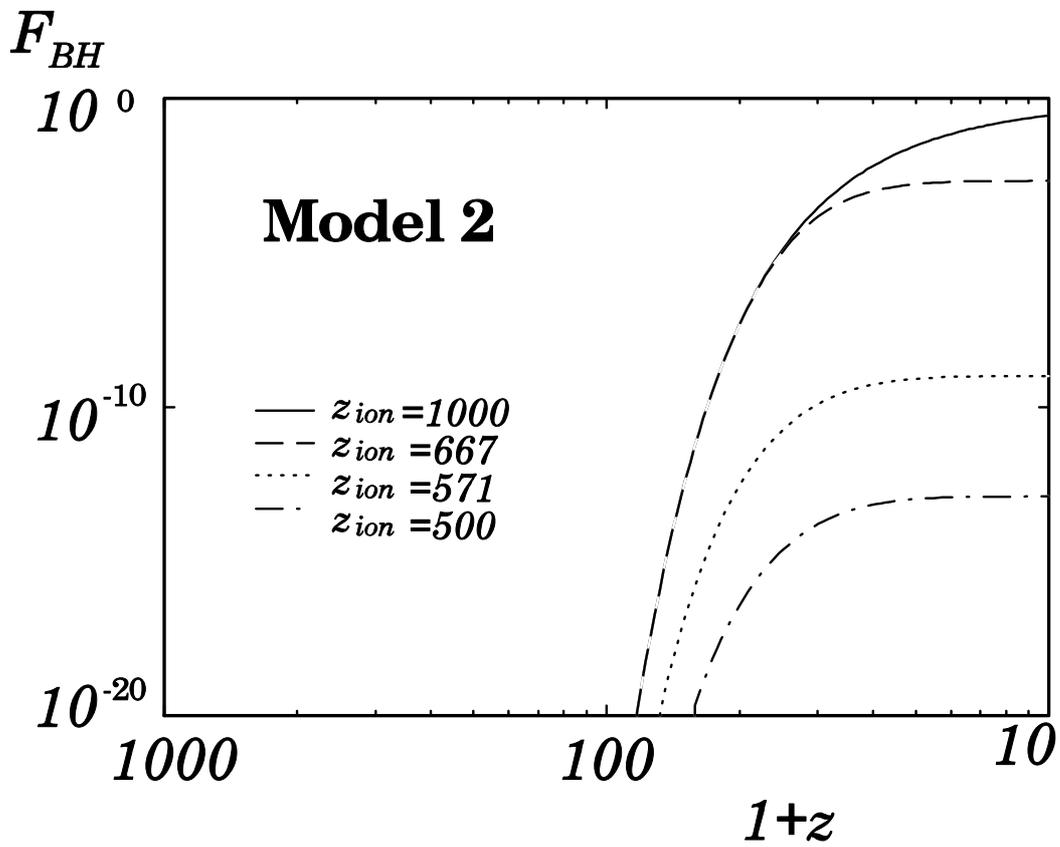
Fig.6b

# Probability Distribution of Primordial Angular Momentum and Formation of Massive Black Holes

HAJIME SUSA, MISAO SASAKI, AND TAKAHIRO TANAKA

*Department of Physics, Kyoto University, Kyoto 606-01, Japan*

## ABSTRACT

We consider the joint probability distribution function for the mass contrast and angular momentum of over-density regions on the protogalactic scale and investigate the formation of massive black holes at redshift $z \gtrsim 10$. We estimate the growth rate of the angular momentum by the linear perturbation theory and the decay rate by the Compton drag and apply the Press-Schechter theory to obtain the formation rate of massive black holes, assuming the full reionization of the universe at $z = z_{ion} \gg 10$. We find the correlation between the mass contrast and angular momentum vanishes in the linear theory. However, application of the Press-Schechter theory introduces a correlation between the mass contrast and angular momentum of bound objects. Using thus obtained probability distribution, we calculate the mass fraction of black holes with $M \sim 10^6 - 10^8 M_\odot$ in the universe. We find that it crucially depends on the reionization epoch $z_{ion}$. Specifically, for the standard CDM power spectrum with the COBE normalization, the condition $z_{ion} \gtrsim 500$ must be satisfied to reproduce the observed number density of QSOs.

# 1. Introduction

Now in the cosmology, one of the least known periods lies between the recombination time and the redshift $z \sim 5$ at which most distant QSOs are observed. This era has given us very little observational information, but is believed to be a stage during which various important events and processes for the structure formation occurred. Among them, one that has recently attracted much attention is the reionization of the universe. If a sufficient number of massive stars had formed soon after recombination, the universe would be reionized at very high redshifts $z \gtrsim 100$.[1] Although we have no firm theoretical ground to estimate the number of such massive stars at the moment, there are various observational reasons which favor the reionization scenario.

Concerning the effect of reionization on the large scale structure, anisotropies of the cosmic microwave background (CMB) at intermediate angular scales ($\sim$ arcminute to degree scales) would be reduced considerably to loosen the constraint on the amplitude of the primordial density perturbation on corresponding scales ($\sim$1 to 100 Mpc).[2] In particular, this may help explaining the existence of large scale velocity fields without contradicting the observed upper bound of the degree scale CMB anisotropy.[3]

Reionization of the matter also strongly affects the formation of the small scale structure of the universe. Right after the recombination, the Jeans mass for baryons is $\sim 10^6 M_\odot$ and in the absence of reionization it gradually decreases as the universe expands. However, if the universe was reionized, there would have been a considerable increase in the Jeans mass. This would mean the interruption of the small scale structure formation. This situation can be avoided if the dominant component of the universe is some cold dark matter (CDM), since the Jean mass for CDM is negligibly small. Then the baryons fall into the CDM potential well to form protogalactic clouds and they may start collapsing once the optical depth becomes sufficiently small and the cooling time is short compared with the free-fall time. Once we assume such a scenario, then a matter of interest is whether those protogalactic clouds would collapse completely to form black holes, since the formation of such massive black holes at high redshifts can explain the existence of QSOs and/or AGNs at $z \sim 5$.

Closely related subjects have been discussed previously by various authors. It is already shown that the number density of QSOs can be explained[4,5] in the context of the Press-Schechter theory,[6,7] which assumes a simple extrapolation of linear theory to the non-linear growth of density perturbations. However, they neglected the effect of angular momentum. As was pointed out by Peebles,[8] the angular momentum gained during the growth of the density perturbation may be large enough to support the cloud against runaway collapse into a black hole.[†] Loeb[13] and Umemura et al.[14] calculated the angular momentum barrier of the density perturbation and showed that the angular momentum of the cloud can be reduced sufficiently to form a black hole by Compton drag, i.e., the inverse Compton scattering with CMB photons, provided that the reionization epoch was early enough. Then what should be done now is a statistical analysis to evaluate the fraction of baryons which turns into black holes, given a primordial density perturbation spectrum.

In this paper, assuming the reionization scenario as mentioned above, we investigate the formation of massive black holes ($M \sim 10^6 - 10^8 M_\odot$) at high redshifts and calculate the mass fraction of those black holes in the universe to test whether such a reionization scenario can explain the observed number density of QSOs. We assume a CDM dominated universe with the Harrison-Zeldovich spectrum[15] and normalize the amplitude by the COBE-DMR result at $10°$.[16] Reionization is assumed to have occurred at once at $z = z_{ion}$, which we regard as a free parameter. The paper is organized as follows. In section 2, we estimate the angular momentum barrier of a cloud by linear theory and calculate the joint probability distribution $P(\vec{J}(r_0), \delta(r_0); t)$ for the angular momentum $\vec{J}(r_0)$ and mass contrast $\delta(r_0)$ of protogalactic clouds. We find there is no correlation between the angular momentum and mass contrast in linear theory. In section 3, we apply a Press-Schechter-like theory to the joint probability distribution function $P(\vec{J}(r_0), \delta(r_0); t)$ to formulate

---

† Although there have been arguments on the proper definition of the angular momentum,[9−12] in this paper, we adopt the one according to Peebles.[8] See section 2 below for further discussion on this point.

– 3 –

a method to calculate the mass fraction of black holes in the universe, taking into account the loss of the angular momentum by the Compton drag. In doing so, a correlation between the angular momentum and mass contrast is introduced. In section 4, we show our results of calculations. Within the context of CDM scenario with the COBE normalization, we find $z_{ion}$ must be greater than 500 to explain the observed number density of QSOs. Finally, section 5 is devoted to summary and discussion.

## 2. Linear theory

In this section, we derive the probability distribution function for the mass contrast $\delta(r_0)$ and the angular momentum $\vec{J}(r_0)$ averaged over a sphere of comoving radius $r_0$ in the linear perturbation theory.

For definiteness, we adopt the standard picture of the very early universe, namely a cosmological model based on the inflationary scenario, and assume $\Omega = 1$. As the epoch we are interested in is the matter dominated era, we set

$$a(t) = \left(\frac{t}{t_0}\right)^{2/3}, \quad \rho(t) = \frac{\rho_0}{a^3}, \tag{2.1}$$

where $\rho_0$ is the cosmic mean density at the present time and $t_0$ is the age of the universe. Furthermore we assume the universe is cold dark matter dominated and baryons contribute only a small fraction of the cosmic density, $\Omega_b \ll 1$. Thus we assume $\delta_b = \delta$ at the stage of our interest.

Following Peebles,[8] we define the angular momentum measured not from the geometrical center of the sphere but from its center of mass. Hence,

$$\begin{aligned}
\delta(r_0) &:= \frac{3}{4\pi r_0^3} \int_{|\boldsymbol{r}|\leq r_0} \delta(t,\boldsymbol{r})d^3r, \\
\vec{J}(r_0) &:= \Omega_b a^4 \int_{|\boldsymbol{r}|\leq r_0} \rho(t)(1+\delta(t,\boldsymbol{r}))(\boldsymbol{r} - \boldsymbol{R}) \times \boldsymbol{u}(t,\boldsymbol{r})d^3r,
\end{aligned} \tag{2.2}$$

where $\boldsymbol{u}(t,\boldsymbol{r})$ is the peculiar velocity field, $\boldsymbol{r}$ represents the comoving coordinates,

and $\boldsymbol{R}$ is the center of mass coornidate of the sphere,

$$\boldsymbol{R} = \frac{3}{4\pi r_0^3} \int_{|\boldsymbol{r}|\leq r_0} \delta(t,\boldsymbol{r})\boldsymbol{r}\, d^3r. \tag{2.3}$$

Note that our definition of the angular momentum is second order in the perturbation amplitude, which is different from the first order expression one obtains in the Lagrangian picture as in some other recent works.[9-12] But provided we consider the angular momentum gain until $\delta$ becomes of order unity, the resulting magnitude of the angular momentum is not so different from each other. Furthermore, in our models discussed in section 3, if the angular momentum loss due to Compton drag is absent, the resulting spin parameter $\lambda := J\sqrt{E}G^{-1}M^{-5/2}$ turns out to be mass-scale independent. Here $E$ denotes the binding energy of the density perturbation. Specifically, we obtain $\lambda \simeq 0.025$ for *Model 1* and $\lambda \simeq 0.08$ for *Model 2* (see Eqs.(3.1) and (3.2) below), while the value $\lambda \simeq 0.05$ has been obtained in numerical simulations.[17] Hence after all, our definition is not a bad approximation for the primodial angular momentum. Note also that, as we are interested in the baryonic component, the angular momentum we have defined is that of baryons.

Under the assumption of the rotation-free velocity field and the Gaussian nature of the primordial density perturbations, which are also naturally predicted in the inflationary universe scenario, the density perturbation $\delta$ and the velocity field $\boldsymbol{u}(t,\boldsymbol{r})$ may be expressed in terms of the Fourier components as

$$\begin{aligned}\delta(t,\boldsymbol{r}) &= \int d\boldsymbol{k}\, \sigma_{\boldsymbol{k}}(t) z_{\boldsymbol{k}} e^{i\boldsymbol{k}\cdot\boldsymbol{r}}, \\ \boldsymbol{u}(t,\boldsymbol{r}) &= \int d\boldsymbol{k}\, \frac{2i}{3}\frac{a}{t}\frac{\boldsymbol{k}}{k^2}\sigma_{\boldsymbol{k}}(t) z_{\boldsymbol{k}} e^{i\boldsymbol{k}\cdot\boldsymbol{r}},\end{aligned} \tag{2.4}$$

where $\sigma_{\boldsymbol{k}}(t)$ is the square root of the density power spectrum which is proportional to the scale factor in the present case, and $z_{\boldsymbol{k}}$ is an independent stochastic variable for each $\boldsymbol{k}$ except that $z_{-\boldsymbol{k}}$ must be identified with $\bar{z}_{\boldsymbol{k}}$ because of the reality of $\delta$. The distribution of $z_{\boldsymbol{k}}$ is normalized as

$$P_z\left(z_{\boldsymbol{k}}\right) dx_{\boldsymbol{k}} dy_{\boldsymbol{k}} = \frac{dx_{\boldsymbol{k}} dy_{\boldsymbol{k}}}{2\pi} \exp\left(-\frac{x_{\boldsymbol{k}}^2 + y_{\boldsymbol{k}}^2}{2}\right), \tag{2.5}$$

where

$$z_{\bm{k}} = \frac{1}{\sqrt{2}}\left(x_{\bm{k}} + iy_{\bm{k}}\right), \quad x_{-\bm{k}} = x_{\bm{k}}, \quad y_{-\bm{k}} = -y_{\bm{k}}. \tag{2.6}$$

In terms of the these Fourier variables, Eq.(2.2) is rewritten as

$$\begin{aligned}\delta(r_0) &= \sum_{\bm{k}} z_{\bm{k}} g_{k}, \\ \vec{J}(r_0) &= \sum_{\bm{k}\bm{k}'} \vec{f}_{\bm{k}\bm{k}'} z_{\bm{k}} z'_{\bm{k}},\end{aligned} \tag{2.7}$$

where

$$\begin{aligned}g_k &:= W_M(kr_0)\sigma_k, \\ \vec{f}_{\bm{k}\bm{k}'} &:= -\frac{8\pi}{45}\Omega_b \rho \frac{(ar_0)^5}{t} \frac{\bm{k}\times\bm{k}'}{k'^2}\sigma_k \sigma_{k'}\left(W_J(|\bm{k}+\bm{k}'|r_0) - W_J(kx_0)W_M(k'r_0)\right) \\ &\quad - (\bm{k}\to\bm{k}',\bm{k}'\to\bm{k}),\end{aligned} \tag{2.8}$$

and $W_M$ and $W_J$ are the Fourier transforms of the top hat window function and its derivative, respectively, which are given by

$$W_M(y) := 3\frac{j_1(y)}{y}, \qquad W_J(y) := 15\frac{j_2(y)}{y^2}. \tag{2.9}$$

Hereafter we drop the argument $r_0$, for notational simplicity. Since the statistical nature of the density distribution is all known, we can calculate the joint probability distribution of the two variables by performing the following integration,

$$\begin{aligned}P_{J\delta}\left(\vec{J},\delta;t\right) = \int &\delta^D\left(\vec{J} - \sum_{\bm{k}\bm{k}'}\vec{f}_{\bm{k}\bm{k}'}z_{\bm{k}}z'_{\bm{k}}\right) \\ &\times \delta^D\left(\delta_{x_0} - \sum_{\bm{k}}z_{\bm{k}}g_{k}\right) \\ &\times P_z\left(z_{\bm{k}}\right)\prod_{\bm{k}}{}' dx_{\bm{k}} dy_{\bm{k}},\end{aligned} \tag{2.10}$$

where the primed product $\prod'_{\bm{k}}$ represents a product over a half of the Fourier space $\bm{k}$ for which $x_{\bm{k}}$ and $y_{\bm{k}}$ are independent, e.g., those which satisfy $k_x \geq 0$. Using the

Fourier transform of Dirac's $\delta$-function, we rewrite Eq.(2.10) as

$$P_{J\delta}\left(\vec{J},\delta;t\right) = \int\int \frac{d^3s}{(2\pi)^3}\frac{dn}{2\pi}\exp\left(in\delta\right)\exp\left(i\vec{s}\cdot\vec{J}\right)$$
$$\times \int\prod_{\bm{k}}{}' \frac{dx_{\bm{k}}dy_{\bm{k}}}{2\pi}\exp\left(-\frac{1}{2}\sum_{\bm{k}\bm{k}'}{}' M_{\bm{k}\bm{k}'}x_{\bm{k}}x'_{\bm{k}}\right) \quad (2.11)$$
$$\times \exp\left(-in\sqrt{2}\sum_{\bm{k}}{}' x_{\bm{k}}g_{\bm{k}}\right)\exp\left(-\frac{1}{2}\sum_{\bm{k}\bm{k}'}{}' \widetilde{M}_{\bm{k}\bm{k}'}y_{\bm{k}}y'_{\bm{k}}\right),$$

where

$$\begin{aligned}
M_{\bm{k}\bm{k}'} &:= \delta_{\bm{k}\bm{k}'} + 2i\vec{f}^{+}_{\bm{k}\bm{k}'}\cdot\vec{s}, \\
\widetilde{M}_{\bm{k}\bm{k}'} &:= \delta_{\bm{k}\bm{k}'} - 2i\vec{f}^{-}_{\bm{k}\bm{k}'}\cdot\vec{s}, \\
\vec{f}^{+}_{\bm{k}\bm{k}'} &:= \vec{f}_{\bm{k}\bm{k}'} + \vec{f}_{\bm{k}-\bm{k}'}, \\
\vec{f}^{-}_{\bm{k}\bm{k}'} &:= \vec{f}_{\bm{k}\bm{k}'} - \vec{f}_{\bm{k}-\bm{k}'},
\end{aligned} \quad (2.12)$$

and $\sum'$ is defined similarly as $\prod'$. After the integration with respect to $dx_{\bm{k}}$, $dy_{\bm{k}}$ and $dn$, we obtain

$$P_{J\delta}\left(\vec{J},\delta;t\right) = \int \frac{d^3s}{(2\pi)^3}\exp\left(i\vec{s}\cdot\vec{J}\right)\left(\det\left(M\widetilde{M}\right)\right)^{-\frac{1}{2}}$$
$$\times \frac{1}{\sqrt{4\pi\sum'_{\bm{k}\bm{k}'} M^{-1}_{\bm{k}\bm{k}'}g_{\bm{k}}g_{\bm{k}'}}}\exp\left(-\frac{\delta^2}{4\sum'_{\bm{k}\bm{k}'} M^{-1}_{\bm{k}\bm{k}'}g_{\bm{k}}g_{\bm{k}'}}\right). \quad (2.13)$$

We see that from this expression the possible correlation between $\delta$ and $\vec{J}$ arises only through the $\vec{s}$-dependence of the last exponential factor,

$$\exp\left(-\frac{\delta^2}{4\sum'_{\bm{k}\bm{k}'} M^{-1}_{\bm{k}\bm{k}'}g_{\bm{k}}g_{\bm{k}'}}\right).$$

Interestingly, this term turns out to be $\vec{s}$-independent. Consequently there exists no correlation between $\delta$ and $\vec{J}$ in the linear perturbation theory. Now let us show

this fact, or that

$$C := M^{-1}_{kk'} g_k g_{k'}, \qquad (2.14)$$

is $\vec{s}$-independent. To show this, we expand $C$ in the power series of $\vec{s}$ as

$$C = \sum_{N=0}^{\infty} C_N, \qquad (2.15)$$

where

$$C_0 = {\sum_{k}}' g_k^2,$$
$$C_N = {\sum_{k,k_1 k_2 \ldots k_{N-1}, k'}}' g_k g_{k'} (-i\vec{s} \cdot \vec{f}^+_{kk_1}) \ldots (-i\vec{s} \cdot \vec{f}^+_{k_{N-1} k'}). \qquad (2.16)$$

¿From the definition of $f^+_{kk'}$ given in Eq.(2.12), we can rewrite the primed sum over the last index $k'$ in $C_N$ to the unprimed sum as

$$C_N = {\sum_{k,k_1 k_2 \ldots k_{N-1}}}' g_k (-i\vec{s} \cdot \vec{f}^+_{kk_1}) \ldots (-i\vec{s} \cdot \vec{f}^+_{k_{N-1} k'}) \sum_{k'} g_{k'} (-i\vec{s} \cdot \vec{f}_{k_{N-1} k'}). \qquad (2.17)$$

Now, since $f_{k_{N-1} k'}$ given in Eq.(2.8) has the form,

$$\vec{f}_{kk'} = k \times k' \left( \mathcal{W}(|k + k'|, k, k') - \mathcal{W}(|k + k'|, k', k) \right), \qquad (2.18)$$

it is easy to see that for each $k$ there exists one and the only one $\widetilde{k}$ whose contribution exactly cancels that from $k$ (see Fig.1). Thus $C_N = 0$ for all $N \geq 1$ and we obtain

$$C = {\sum_{k}}' g_k^2 = \frac{1}{2} \sum_{k} g_k^2. \qquad (2.19)$$

As a result, the joint probability distribution factorizes:

$$P_{J\delta}(\vec{J}, \delta; t) = P_\delta(\delta; t) \cdot P_J(\vec{J}; t), \qquad (2.20)$$

where

$$P_\delta(\delta;t) = \frac{1}{\sqrt{2\pi\langle\delta^2\rangle}}\exp\left(-\frac{\delta^2}{2\langle\delta^2\rangle}\right); \quad \langle\delta^2\rangle = \sum_{\mathbf{k}} g_k^2,$$

$$P_J(\vec{J};t) = \int \frac{d^3s}{(2\pi)^3}\exp\left(i\vec{s}\cdot\vec{J}\right)\left(\det\left(M\widetilde{M}\right)\right)^{-\frac{1}{2}}.$$

(2.21)

In contrast to the apparent Gaussian nature of $P_\delta$, the explicit form of $P_J$ is hard to obtain. One case that can be analytically calculable is the limit of large $|\vec{J}|$, for which we may approximate the determinant factor by exponetiating it, $(\det M\widetilde{M})^{-1/2} = \exp\left[-\frac{1}{2}\operatorname{Tr}\ln(M\widetilde{M})\right]$, amd expand the exponent in powers of $\vec{s}$ to $O(\vec{s}^2)$. The resulting form of the probability distribution is Gaussian,

$$P_J(\vec{J};t)d^3J = \frac{1}{\left(2\pi\langle\vec{J}^2\rangle/3\right)^{3/2}}\exp\left(-\frac{\vec{J}^2}{2\langle\vec{J}^2\rangle/3}\right)d^3J; \quad \langle\vec{J}^2\rangle = 2\sum_{\mathbf{k}\mathbf{k}'}\vec{f}_{\mathbf{k}\mathbf{k}'}^2.$$

(2.22)

Although there is no justification for this to hold when $|\vec{J}|$ is small, hereafter, we assume the above Gaussian form for $P_J$.†

## 3. The collapsed mass fraction formula

Now we have the approximate factorized joint probability distribution, valid at the linear stage. Its temporal behavior is completely determined by that of $\langle\delta^2\rangle$ and $\langle\vec{J}^2\rangle$. For $\langle\delta^2\rangle$, we simply extrapolate the linear evolution to the non-linear stage, but apply the Press-Schechter theory to interpret the resulting probability distribution. As for $\langle\vec{J}^2\rangle$, we estimate its growth rate by the linear theory, while we estimate its decay rate by including the effect of Compton drag. Specifically, we formulate the evolution of $\langle\vec{J}^2\rangle$ as follows.

First, we consider two models for the growth rate of the angular momentum.

---

† A preliminary numerical evaluation of $P_J$ suggests that its magnitude is enhanced at small $|\vec{J}|$ relative to the Gaussian case, which may affect our analysis below to certain extent.

*Model 1*

$$\left(\frac{d\langle \vec{J}^2 \rangle}{dt}\right)_+ = \begin{cases} \frac{10}{3}\langle \vec{J}^2 \rangle/t & \text{for } \delta \leq \delta_m, \\ 0 & \text{for } \delta > \delta_m, \end{cases} \qquad (3.1)$$

where $\delta_m = 1.06$ which corresponds to the maximum expansion epoch of the spherical dust collapse (Fig.2). Thus the linear theory is applied until this epoch in this model.

*Model 2*

$$\left(\frac{d\langle \vec{J}^2 \rangle}{dt}\right)_+ = \begin{cases} \frac{10}{3}\langle \vec{J}^2 \rangle/t & \text{for } \delta \leq \delta_c, \\ 0 & \text{for } \delta > \delta_c, \end{cases} \qquad (3.2)$$

where $\delta_c=1.69$ which corresponds to the collapse epoch of the spherical dust collapse (Fig.2). Hence, in this model, the linear growth rate of the angular momentum is extrapolated until the last moment of collapse if the cloud were a homogeneous dust sphere.

On may regard the above two models as representing two extreme cases. In the former model, $\delta = \delta_m$ is roughly the validity limit of the linear theory. So, in this model, the growth of the angular momentum in the non-linear phase is ignored. Hence we expect it to give the final angular momentum less than the real value. At the non-linear stage, the growth rate will decrease due to the decrease in the quadrupole moment of the cloud. So, the latter model, in which the linear growth rate is extrapolated to the final stage of collapse, is expected to give the final angular momentum larger than the real value.

Next, we consider the decay of the angular momentum due to Compton drag. We assume a sudden ionization history as

$$\chi_e(z) = \begin{cases} 10^{-4} & \text{for } z > z_{ion}, \\ 1 & \text{for } z < z_{ion}, \end{cases}$$

where $\chi_e$ is the ionization rate. With this ionization history, we model the decay rate of the angular momentum as

$$\left(\frac{d\left\langle \vec{J}^2 \right\rangle}{dt}\right)_- = \begin{cases} -2\alpha_0 (1+z)^4 \left\langle \vec{J}^2 \right\rangle & \text{for } \delta \leq 1.69, \\ 0 & \text{for } \delta > 1.69, \end{cases} \quad (3.3)$$

where

$$\alpha_0(z) = \gamma_0 \chi_e(z) = \begin{cases} \gamma_0 \times 10^{-4} =: \alpha_1 & \text{for } z > z_{ion}, \\ \gamma_0 \times 1 =: \alpha_2 & \text{for } z < z_{ion}, \end{cases} \quad (3.4)$$

$$\gamma_0 = \frac{4\sigma_T \epsilon_0}{3\mu m_p c}.$$

Here $\sigma_T$ is the Thomson scattering cross section, $\epsilon_0$ is the energy density of CMB today, $m_p$ is the proton mass, $c$ is the speed of light, and $\mu$ is the mean molecular weight which we assume to be 1, respectively. We cut the Compton drag at the collapse epoch, since the photon trapping occurs almost at this epoch (Loeb 1993).

Now we solve the differential equation,

$$\left(\frac{d\left\langle \vec{J}^2 \right\rangle}{dt}\right) = \left(\frac{d\left\langle \vec{J}^2 \right\rangle}{dt}\right)_+ + \left(\frac{d\left\langle \vec{J}^2 \right\rangle}{dt}\right)_-. \quad (3.5)$$

This can be analytically solved. We define the amplitude of $\delta$ which characterize the reionization epoch,

$$\delta_{ion} := \delta \frac{1+z}{1+z_{ion}}. \quad (3.6)$$

Then the solution is characterized by the value of $\delta_{ion}$. We first show the results for *Model 1*.

Case 1A; $\delta_{ion} \leq \delta_m$.

In this case, the evolution of $\left\langle \vec{J}^2 \right\rangle$ is divided into the following four stages:

(i) $\delta \leq \delta_{ion}$;
$$\frac{\left\langle \vec{J}^2 \right\rangle}{\left\langle \vec{J}_i^2 \right\rangle} = \left(\frac{1+z_i}{1+z}\right)^5 \exp\left[-\frac{6\alpha_1 t_0}{5}\left((1+z_i)^{5/2} - (1+z)^{5/2}\right)\right],$$

(ii) $\delta_{ion} \leq \delta \leq \delta_m$;
$$\frac{\left\langle \vec{J}^2 \right\rangle}{\left\langle \vec{J}_{ion}^2 \right\rangle} = \left(\frac{1+z_{ion}}{1+z}\right)^5 \exp\left[-\frac{6\alpha_2 t_0}{5}\left((1+z_{ion})^{5/2} - (1+z)^{5/2}\right)\right],$$

(iii) $\delta_m < \delta \leq \delta_c$;
$$\frac{\left\langle \vec{J}^2 \right\rangle}{\left\langle \vec{J}_m^2 \right\rangle} = \exp\left[-\frac{6\alpha_2 t_0}{5}\left((1+z_m)^{5/2} - (1+z)^{5/2}\right)\right],$$

(iv) $\delta_c < \delta$;
$$\frac{\left\langle \vec{J}^2 \right\rangle}{\left\langle \vec{J}_m^2 \right\rangle} = \text{const.}.$$

(3.7)

Case 1B; $\delta_m \leq \delta_{ion} \leq \delta_c$.

We also have four different stages of the evolution in this case:

(i) $\delta \leq \delta_m$ ;

$$\frac{\left\langle \vec{J}^2 \right\rangle}{\left\langle \vec{J}_i^2 \right\rangle} = \left(\frac{1+z_i}{1+z}\right)^5 \exp\left[-\frac{6\alpha_1 t_0}{5}\left((1+z_i)^{5/2} - (1+z)^{5/2}\right)\right],$$

(ii) $\delta_m < \delta \leq \delta_{ion}$ ;

$$\frac{\left\langle \vec{J}^2 \right\rangle}{\left\langle \vec{J}_m^2 \right\rangle} = \exp\left[-\frac{6\alpha_1 t_0}{5}\left((1+z_m)^{5/2} - (1+z)^{5/2}\right)\right],$$

(3.8)

(iii) $\delta_{ion} < \delta \leq \delta_c$ ;

$$\frac{\left\langle \vec{J}^2 \right\rangle}{\left\langle \vec{J}_{ion}^2 \right\rangle} = \exp\left[-\frac{6\alpha_2 t_0}{5}\left((1+z_{ion})^{5/2} - (1+z)^{5/2}\right)\right],$$

(iv) $\delta_c < \delta$ ;

$$\frac{\left\langle \vec{J}^2 \right\rangle}{\left\langle \vec{J}_{ion}^2 \right\rangle} = \text{const.}.$$

Case 1C; $\delta_c \leq \delta_{ion}$ .

In this case, the evolution of $\left\langle \vec{J}^2 \right\rangle$ is divided into three stages:

(i) $\delta \leq \delta_m$ ;

$$\frac{\left\langle \vec{J}^2 \right\rangle}{\left\langle \vec{J}_i^2 \right\rangle} = \left(\frac{1+z_i}{1+z}\right)^5 \exp\left[-\frac{6\alpha_1 t_0}{5}\left((1+z_i)^{5/2} - (1+z)^{5/2}\right)\right],$$

(ii) $\delta_m < \delta \leq \delta_c$ ;

$$\frac{\left\langle \vec{J}^2 \right\rangle}{\left\langle \vec{J}_m^2 \right\rangle} = \exp\left[-\frac{6\alpha_1 t_0}{5}\left((1+z_m)^{5/2} - (1+z)^{5/2}\right)\right],$$

(3.9)

(iii) $\delta_c < \delta$ ;

$$\frac{\left\langle \vec{J}^2 \right\rangle}{\left\langle \vec{J}_m^2 \right\rangle} = \text{const.}.$$

In the above, the subscript "$i$" denotes the initial epoch which we take to be the decoupling time, "$m$" the epoch at which $\delta = \delta_m$ and "$ion$" at $z = z_{ion}$.

As for *Model 2*, the evolutionary behavior becomes a bit simpler, since the epoch $\delta = \delta_m$ plays no role any more.

Case 2A; $\delta_{ion} \leq \delta_c$.

The evolution of $\left\langle \vec{J}^2 \right\rangle$ is divided into three stages:

(i) $\delta \leq \delta_{ion}$;
$$\frac{\left\langle \vec{J}^2 \right\rangle}{\left\langle \vec{J}_i^2 \right\rangle} = \left(\frac{1+z_i}{1+z}\right)^5 \exp\left[-\frac{6\alpha_1 t_0}{5}\left((1+z_i)^{5/2} - (1+z)^{5/2}\right)\right],$$

(ii) $\delta_{ion} \leq \delta \leq \delta_c$;
$$\frac{\left\langle \vec{J}^2 \right\rangle}{\left\langle \vec{J}_{ion}^2 \right\rangle} = \left(\frac{1+z_{ion}}{1+z}\right)^5 \exp\left[-\frac{6\alpha_2 t_0}{5}\left((1+z_{ion})^{5/2} - (1+z)^{5/2}\right)\right],$$

(iii) $\delta_c < \delta$;
$$\frac{\left\langle \vec{J}^2 \right\rangle}{\left\langle \vec{J}_c^2 \right\rangle} = \text{const.}.$$

(3.10)

Case 2B; $\delta_m \leq \delta_{ion} \leq \delta_c$.

The evolution of $\left\langle \vec{J}^2 \right\rangle$ is divided into two stages:

(i) $\delta \leq \delta_c$;
$$\frac{\left\langle \vec{J}^2 \right\rangle}{\left\langle \vec{J}_i^2 \right\rangle} = \left(\frac{1+z_i}{1+z}\right)^5 \exp\left[-\frac{6\alpha_1 t_0}{5}\left((1+z_i)^{5/2} - (1+z)^{5/2}\right)\right],$$

(3.11)

(ii) $\delta_c < \delta$;
$$\frac{\left\langle \vec{J}^2 \right\rangle}{\left\langle \vec{J}_c^2 \right\rangle} = \text{const.},$$

where the subscript "c" denotes the epoch $\delta = \delta_c$.

From the above results, the final value of the angular momentum after a clould is ready to collapse adiabatically can be estimated. For *Model 1*, we obtain

Cases 1A and 1B; $\delta_{ion} \leq \delta_c$ :

$$\frac{\left\langle \vec{J}_f^2 \right\rangle}{\left\langle \vec{J}_i^2 \right\rangle} = 2^{-10/3} \left( \frac{1+z_i}{1+z_c} \right)^5$$
$$\times \exp\left[ -\frac{6t_0}{5} \left( \alpha_1 \left( (1+z_i)^{5/2} - (1+z_{ion})^{5/2} \right) \right. \right. \tag{3.12}$$
$$\left. \left. + \alpha_2 \left( (1+z_{ion})^{5/2} - (1+z_c)^{5/2} \right) \right) \right],$$

Case 1C; $\delta_c \leq \delta_{ion}$ :

$$\frac{\left\langle \vec{J}_f^2 \right\rangle}{\left\langle \vec{J}_i^2 \right\rangle} = 2^{-10/3} \left( \frac{1+z_i}{1+z_c} \right)^5$$
$$\times \exp\left[ -\frac{6\alpha_1 t_0}{5} \left( (1+z_i)^{5/2} - (1+z_c)^{5/2} \right) \right], \tag{3.13}$$

and for *Model 2*, we obtain

Case 2A; $\delta_{ion} \leq \delta_c$ :

$$\frac{\left\langle \vec{J}_f^2 \right\rangle}{\left\langle \vec{J}_i^2 \right\rangle} = \left( \frac{1+z_i}{1+z_c} \right)^5$$
$$\times \exp\left[ -\frac{6t_0}{5} \left( \alpha_1 \left( (1+z_i)^{5/2} - (1+z_{ion})^{5/2} \right) \right. \right. \tag{3.14}$$
$$\left. \left. + \alpha_2 \left( (1+z_{ion})^{5/2} - (1+z_c)^{5/2} \right) \right) \right],$$

Case 2B; $\delta_c \leq \delta_{ion}$ :

$$\frac{\left\langle \vec{J}_f^2 \right\rangle}{\left\langle \vec{J}_i^2 \right\rangle} = \left(\frac{1+z_i}{1+z_c}\right)^5 \tag{3.15}$$
$$\times \exp\left[-\frac{6\alpha_1 t_0}{5}\left((1+z_i)^{5/2} - (1+z_c)^{5/2}\right)\right],$$

where $\left\langle \vec{J}_f^2 \right\rangle$ denotes the final mean square value of the angular momentum. We see that the final rms angular momentum for *Model 2* is larger than that for *Model 1* by the factor $2^{5/3} = 3.17$.

These solutions are plotted in Fig.3 for a variety of $z_{ion}$. The amplitude of the density perturbation is set to the rms value estimated from the standard CDM model at scale $M = 10^{10} M_\odot$. As easily seen from the figure, the final value of the angular momentum depends extremely sensitively on $z_{ion}$. As we have seen in the above, it also depends on the model of $d\left\langle \vec{J}^2 \right\rangle_+/dt$ but not much compared to the $z_{ion}$-dependence.

Having obtained the evolutionary model for $\left\langle \vec{J}^2 \right\rangle$, we now have that of the joint probability distribution $P_{J\delta}(\vec{J}, \delta; t)$. Note that it does not factorize any more, due to the $\delta$-dependence of $\left\langle \vec{J}^2 \right\rangle$ we have introduced. Using thus obtained probability distribution, we formulate a method to calculate the collasped mass fraction. As for $\delta$, we use the Press-Schechter theory to estimate the collapsed fraction. As for $\vec{J}$, we sum up the fractions which satisfy $|\vec{J}| \leq J_{BH}$, where $J_{BH}$ is the maximum angular momentum which a Kerr black hole can hold,

$$J_{BH} = \frac{GM_b^2}{c}, \tag{3.16}$$

where $M_b$ is the baryon mass contained in the mass scale $M$.

Thus the mass fraction of baryons which collapsed into massive black holes on scale $M$ by the time $t$ is

$$F_{BH}(x_0;t) = \int_{\delta_c}^{\infty} d\delta \int_{|\vec{J}|\leq J_{BH}} d^3\vec{J} P_{J\delta}\left(\vec{J},\delta;t\right)$$

$$= \int_{\delta_c}^{\infty} d\delta \frac{1}{\sqrt{2\pi\sum_{\boldsymbol{k}}g_{\boldsymbol{k}}^2}} \exp\left(-\frac{\delta^2}{2\sum_{\boldsymbol{k}}g_{\boldsymbol{k}}^2}\right) \qquad (3.17)$$

$$\times \int_0^{J_{BH}} J^2 dJ \frac{4\pi}{\sqrt{2\pi\langle\vec{J}^2\rangle}^3} \exp\left(-\frac{J^2}{2\langle\vec{J}^2\rangle}\right).$$

## 4. Analysis and results

In this section, we apply the formulae derived in the previous section to a cosmological model with the standard CDM density perturbation spectrum and present the results. We assume $\Omega_b = 0.05$ and $h = 0.5$.

Let us first consider the scale dependence of $\langle\vec{J}_f^2\rangle$. Since $\langle\vec{J}_f^2\rangle \propto \langle\vec{J}_i^2\rangle$, it is the same as that of $\langle\vec{J}_i^2\rangle$, which is evaluated as

$$\langle\vec{J}_i^2\rangle^{1/2} \simeq \frac{8\pi}{45}\Omega_b \left[\frac{\rho\,(ar_0)^5}{t}\langle\delta^2\rangle\right]_{t=t_i}$$
$$= \frac{2}{15}\left(\frac{3}{4\pi}\right)^{2/3} \Omega_b \frac{\delta_c^2(1+z_c)^2}{(1+z_i)^{5/2}\rho_0^{2/3} t_0} M^{5/3}. \qquad (4.1)$$

In the above, the value of $\langle\delta(t_i)^2\rangle$ (or $z_c$) depends on the scale $M$ in general, but within the range of our interest, $10^8 M_\odot \lesssim M \lesssim 10^{10} M_\odot$ (corresponding to the black hole mass of $10^6 M_\odot \lesssim M \lesssim 10^8 M_\odot$), its dependence is very weak for the CDM spectrum. Hence we may conclude that

$$\langle\vec{J}_f^2\rangle^{1/2} \propto M^{5/3}. \qquad (4.2)$$

On the other hand, form Eq.(3.16), the maximum angular momentum of a black

hole is

$$J_{BH} = \frac{GM^2}{c}\Omega_b^2 . \qquad (4.3)$$

Comparing this scale dependence of $J_{BH}$ with that of $\left\langle \vec{J}_f^2 \right\rangle^{1/2}$, we find that there exists a critical scale $M_{cr}$ at which $\left\langle \vec{J}_f^2 \right\rangle = J_{BH}^2$ and we have $\left\langle \vec{J}_f^2 \right\rangle < J_{BH}^2$ for $M > M_{cr}$. Hence density perturbations with the average amplitude can collapse to a black hole for $M > M_{cr}$. This critical scale is calculated to be

$$\begin{aligned} M_{cr1} = &\frac{6}{32 \cdot 125}\Omega_b^{-3}\delta_c^{\ 6}(1+z_c)^{-3/2}\frac{c^3 t_0}{G} \\ &\times \exp\left[-\frac{9t_0}{5}\left(\alpha_1\left((1+z_i)^{5/2}-(1+z_{ion})^{5/2}\right)\right.\right. \\ &\left.\left.+\alpha_2\left((1+z_{ion})^{5/2}-(1+z_c)^{5/2}\right)\right)\right], \end{aligned} \qquad (4.4)$$

for *Model 1*, and $M_{cr2} = 32M_{cr1}$ for *Model 2*. They are plotted against $z_{ion}$ in Fig.4.

The result for the black hole mass fraction $F_{BH}$ is plotted in Fig.5. As was the case of Fig.3, the reionization epoch $z_{ion}$ plays the most important role in determination of the final black hole mass fraction. For $z_{ion} \gtrsim 700$, more than 10% of all the baryons will collapse into massive black holes. But for $z_{ion} \lesssim 500$, only an infinitesimal fraction will collapse into black holes. This result is almost independent of the model as well as of the mass scale.

## 5. Summary

We have calculated the joint probability distribution for the mean density contrast $\delta$ and the angular momentum $\vec{J}$ of a sphere from the Gaussian distribution of the density perturbation. We have found there is no correlation between the two variables in the linear regime. Then we have applied the Press-Schechter formalism to the density contrast and taken account of the Compton drag effect on the evolution of the angular momentum. By this procedure, we have introduced certain degrees of correlation betweeen $\delta$ and $\vec{J}$.

Using thus calculated probability distribution, we have evaluated the mass fraction of bound objects and distribution of their angular momentum. Then comparing this final angular momentum with the maximum possible one for a black hole, we have estimated the fraction which collapsed into black holes. We have found that the result is extremely sensitive to the epoch of reionization which we treated as a parameter. More quantitatively, compared with the observed number density of QSOs at redshift $z \sim 2$, the reionization epoch ($z_{ion}$) must satisfy at least $z_{ion} \gtrsim 500$ (Fig.5). We remark that this result is rather insensitive to the amplitude of density perturbations, because the acceptable range of reionization epoch is constrained by the Compton drag time scale.

Of course, the present analysis is only a preliminary attempt to understand the history of the universe during the epoch $1000 \lesssim z \lesssim 10$. For example, one future direction we should take is to perform a more self-consistent treatment of the reionization and the evolution of density perturbations. Also, it is necessary to calculate the probability distribution of the angular momentum in more detail and examine how much it differs from the Gaussian form, which we have assumed in this paper.

## ACKNOWLEDGEMENTS

We would like to thank H. Sato, and R. Nishi for useful discussions. This work was supported in part by the Monbusho Grant-in-Aid for Scientific Research Fund, Nos.2010, 3077 and 05640342.

# REFERENCES

1. M. Kawasaki and M. Fukugita, preprint (1993).

2. P.J.E. Peebles, Astrophys. J. **315** (1987), L73.

3. K.M. Gorski, Astrophys. J. **398** (1992), L5.

4. G. Efstathiou and M.J. Rees, Mon. Not. R. astro. Soc. **230** (1988), 5.

5. M.G. Haehnelt and M.J. Rees, preprint (1993)

6. W.H. Press and P. Schechter, Astrophys. J. **187** (1974), 425.

7. Y. Suto, Prog. Theor. Phys. **90** (1993), 1173.

8. P.J.E. Peebles, Astrophys. J. **155** (1969), 393.

9. A.G. Droshkevich, Astrofisika **6** (1970), 581.

10. S.D.M. White, Astrophys. J. **286** (1984), 38.

11. B.S. Ryden, Astrophys. J. **329** (1988), 589.

12. T. Quinn and J. Binney, Mon. Not. R. astro. Soc. **255** (1992), 729.

13. A. Loeb, Astrophys. J. **403** (1993), 542.

14. M. Umemura, A. Loeb and E.L. Turner, Submitted to Astrophys. J. Letters (1993).

15. J.M. Bardeen, J.R. Bond, N. Kaiser and A.S. Szalay, Astrophys. J. **304** (1986), 15.

16. G.F. Smoot et al., Astrophys. J. **396** (1992), L1.

17. J. Barnes and G. Efstathiou, Astrophys. J. **319** (1987), 575.

# FIGURE CAPTIONS

1. The existence of the wave vector $\widetilde{\boldsymbol{k}}'$ corresponding to any wave vector $\boldsymbol{k}'$, for a fixed $\boldsymbol{k}_{N-1}$, is shown. The contribution of the vector $\widetilde{\boldsymbol{k}}'$ completely cancels that of $\boldsymbol{k}'$ in the sum of Eq.(2.17).

2. A schematic picture of the radius of a collapsing homogeneous dust sphere as a function of the corresponding density contrast $\delta$ in linear theory. The radius reaches its maximum at the epoch when the linear theory gives $\delta = 1.06$ and collapses to zero when $\delta = 1.69$.

3. The evolution of the mean square of the angular momentum on scale $M = 10^{10} M_\odot$. Fig.3a and Fig.3b correspond to the two models in the text. In both figures, different curves correspond to different epochs of the reionization $z_{ion}$ as indicated.

4. The critical mass scale, below which a spherical overdensity region of of the rms perturbation amplitude can collapse to form a black hole, is plotted against the reionization epoch.

5. The mass fraction of baryons which collapsed into massive black holes. The mass scale is $M = 10^{10} M_\odot$, and we have adopted the CDM spectrum with $h = 0.5$ and $\Omega = 1$, normalized with the COBE DMR data at $10°$.

6. The same as Fig.5 but for the mass scale $M = 10^8 M_\odot$.